\documentclass[english]{nato}
\usepackage[latin1]{inputenc}

\pdfoutput=1
\ifx\pdfoptionpdfminorversion\undefined
\newcount\pdfoptionpdfminorversion
\fi

\usepackage{fancyhdr}
\usepackage{graphicx}
\usepackage{amssymb}

\usepackage{txfonts}

\usepackage{sprrefn}

\newdisplay{guess}{Conjecture}

\usepackage{babel}

\begin{document}
\booktitle{Submitted to the proceedings of the NATO Advanced Research Workshop 'Recent Advances in Nonlinear Dynamics and Complex System Physics', Tashkent, Uzbekistan, 2008.}

\firstpage{1}

\begin{opening}

\title{Optomechanics}

\author{Bj\"orn Kubala, Max Ludwig, and Florian Marquardt}

\runningauthor{Bj\"orn Kubala, Max Ludwig, and Florian Marquardt}
\runningtitle{Optomechanics} \institute{Department of Physics,
Arnold Sommerfeld Center for Theoretical Physics, and Center for NanoScience,
Ludwig-Maximilians-Universit\"at, Theresienstrasse 37, 80333 Munich,
Germany}
\begin{abstract}
We review recent progress in the field of optomechanics, where
one studies the effects of radiation on mechanical motion. The paradigmatic
example is an optical cavity with a movable mirror, where the radiation
pressure can induce cooling, amplification and nonlinear dynamics
of the mirror.
\end{abstract}

\end{opening}

\newcommand{\prl}{Phys. Rev. Lett.}

\section{Introduction}

Optomechanics is an emerging research topic that is concerned with
mechanical effects caused by light, particularly in connection with
micro- and nanomechanical structures that are deflected by radiation
pressure. Thoughts about the mechanical effects of light can be traced
back as far as Johannes Kepler. Observing the tails of comets always
pointing away from the sun, he speculated that this might be due to
the force exerted by the solar radiation. Ever since the first measurements
of such radiation forces more than 100 years ago, optomechanical effects
have been observed in various areas of physics and engineering: Spacecraft
with solar sails are indeed being developed, radiation forces are
setting fundamental limits for the precision of laser interferometers
used in detecting gravitational waves, and these forces are also used
to manipulate cold atoms. A recent addition is the use of optomechanical
forces to drive, cool and read out micro- and nanomechanical devices
(see a recent review in \cite{2008_KippenbergVahala_ScienceReview},
and other recent developments in \cite{2008_FM_NaturePhysNewsAndViews}).
To reach the ground state of a mechanical oscillator with a frequency
of $100$~MHz, it would have to be cooled down to about $1$~mK.
Achieving such ground state cooling would {}``put back mechanics
into quantum mechanics'' \cite{2005_SchwabRoukes_PhysicsToday},
and quantum effects would become observable in a massive object consisting
of roughly $10^{15}$ atoms. 

This brief review is organized as follows. In Sec.~\ref{sec:The-basic-optomechanical}
we introduce the basic setup, an optical cavity, driven by a laser
with one mirror placed on an oscillating cantilever. We explain the
classical effects of retarded radiation forces. Similar physics was
investigated in a variety of other system, like driven LC-circuits
coupled to cantilevers \cite{brown:137205} or single-electron transistors
and microwave cavities coupled to nanobeams \cite{2006_Naik_CoolingNanomechResonator,2007_Rodrigues_DynInstabilitiesSSET,2008_Regal_MicrowaveCavity}.
Light-induced forces can not only cool the cantilever, but can also
enhance the mechanical motion leading to an instability. In Sec.~\ref{sec:Nonlinear-classical-dynamic}
we show how one can derive an intricate attractor diagram for the
resulting self-induced oscillations \cite{2006_FM_DynamicalMultistability},
which have also been seen in experiment. Section~\ref{sec:Quantum-theory-of}
is devoted to a quantum description of the coupled cavity-cantilever
system \cite{2008_ML_OptomechInstab}. A new optomechanical setup
\cite{2008_Jayich_NJP,Thompson_2008}, which aims at Fock state detection,
is discussed in Sec.~\ref{sec:Towards-Fock-state-detection}.

\section{The basic optomechanical setup\label{sec:The-basic-optomechanical}}

\begin{figure}

\includegraphics[width=1\columnwidth]{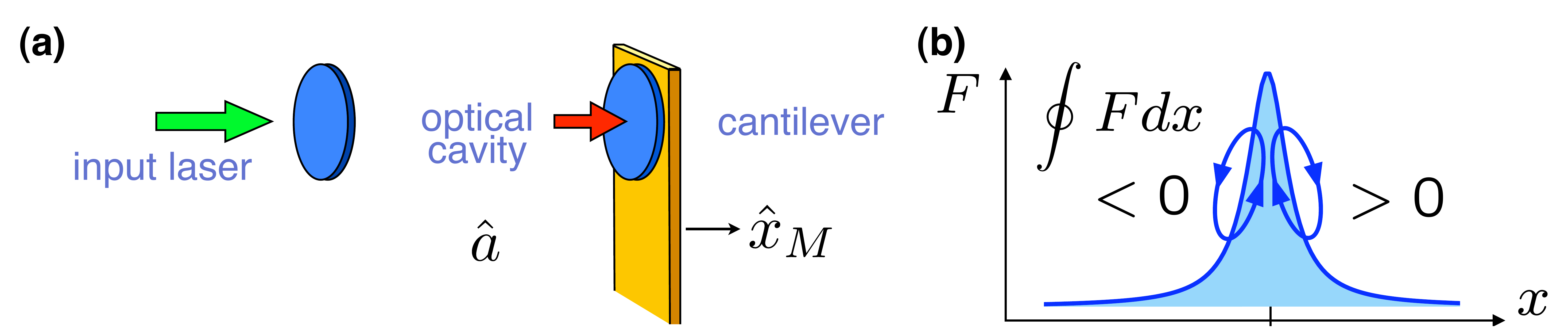}

\caption{ (a) The standard setup of optomechanics.
(b) The dependence of the radiation pressure force (circulating intensity)
on the cantilever position.}
\label{fig:The-standard-setup}
\end{figure}

The standard setup of optomechanics is shown in Fig.~\ref{fig:The-standard-setup}.
It consists of an optical cavity driven by a laser impinging on the
cavity through a fixed mirror. The other mirror of the cavity is movable.
For example, it may be attached to a micro-cantilever as used in atomic
force spectroscopy. In such a setup the mechanical effects of light
are enhanced, as the light field is resonantly increased in the cavity
and each photon will transfer momentum to the mirror in each of the
numerous reflections it undergoes, until finally leaving the cavity.

The coupled cavity-cantilever system is described by a Hamiltonian
of the form\begin{equation}
\hat{H}_{\mathrm{cav+cant}}=\hbar\left(\omega_{\mathrm{cav}}-g\frac{\hat{x}_{M}}{x_{\mathrm{ZPF}}}\right)\hat{a}^{\dagger}\hat{a}+\hbar\omega_{M}\hat{c}^{\dagger}\hat{c}\,.\label{eq:Hamiltonian}\end{equation}
Additional terms in the Hamiltonian describe the driving of the cavity
by the laser beam, decay of photons out of the cavity and the mechanical
damping of the cantilever. Here, $\omega_{M}$ denotes the oscillation
frequency of a mechanical oscillator, whose displacement can be expressed
as $\hat{x}_{M}=(\hat{c}^{\dagger}+\hat{c})x_{\mathrm{ZPF}}$ in terms
of ladder operators and the oscillator's zero point fluctuations $x_{{\rm ZPF}}=(\hbar/2m\omega_{M})^{-1/2}$.
The optical cavity, described by operators $\hat{a}^{\dagger}$ and
$\hat{a}$, has a resonance frequency $\omega_{\mathrm{cav}}$ if
the cantilever is fixed at position $x_{M}=0$.

The coupling term $\propto\hat{x}_{M}\hat{a}^{\dagger}\hat{a}$ with
a strength depending on the coupling constant $g$ can be understood
by two equivalent ways of reasoning: The radiation pressure force
should give rise to a term of the form $-\hat{F}_{\mathrm{rad}}\hat{x}_{M}=-\frac{c}{L}\hat{a}^{\dagger}\hat{a}\,\hbar k_{\mathrm{cav}}\,\hat{x}_{M}$,
which leads to (\ref{eq:Hamiltonian}) with $g=\omega_{\mathrm{cav}}x_{\mathrm{ZPF}}/L$.
Alternatively, we can understand the same term as stemming from the
dependence of the cavity's resonance frequency on the cavity length,
$L+x_{M}$, given by $d\omega_{\mathrm{res}}/dx_{M}=-\omega_{\mathrm{cav}}x_{M}/L$.

Two crucial new ingredients are added to the physics of radiation
pressure by considering a cavity setup. First, the radiation pressure
becomes strongly position dependent due to its proportionality to
the total light intensity in the cavity $\propto\hat{a}^{\dagger}\hat{a}$.
The light intensity shows resonances when the cavity length $L+x_{M}$
is varied. Their full width at half maximum (FWHM) depends on the
decay time $\kappa^{-1}$ of the cavity, $x_{\mathrm{FWHM}}=\kappa L/\omega_{\mathrm{cav}}\,.$The
resulting dependence of the radiation pressure force on the cantilever
position in the stationary state is sketched in Fig.~\ref{fig:The-standard-setup}.
Secondly, the decay time $\kappa^{-1}$ introduces a delay between
the mirror motion and the response of the light intensity. 

To understand the effects of such a retarded response of the radiation
pressure force, let us consider a cantilever at a position $x_{M}>0$
to the right of the resonance (see Fig.~\ref{fig:The-standard-setup})
moving towards the resonance position, $\dot{x}_{M}<0$. We consider
small delay times and small excursions of the cantilever only. Moving
leftwards the cantilever acts against the radiation pressure, which
grows as the cantilever moves closer to resonance and the light intensity
in the cavity increases. This increase, however, lags behind the movement
of the cantilever, so that at any instance the force acting on the
cantilever is smaller than its stationary value at the same position
would be (see Fig.~\ref{fig:The-standard-setup}). Moving into the
opposite, positive direction the delayed decrease of the intensity
leads to an accelerating force on the cantilever, larger than the
stationary one. Overall, there is a net input of work into the mechanical
motion during one oscillation, given by the enclosed area in the force-position
diagram in Fig.~\ref{fig:The-standard-setup}. Thus, for $x_{M}>0$
(where the laser light is blue detuned with respect to the cavity
resonance) the cantilever motion gets enhanced, while for $x_{M}<0$
the same physics causes an additional damping. In the next section,
we will extend these qualitative statements to a detailed description
of the classical dynamics of the coupled cavity-cantilever system.

Retarded radiation forces were first investigated in pioneering studies
by Braginsky, both experimentally and theoretically \cite{1967_BraginskyManukin_PonderomotiveEffectsEMRadiation,1970_Braginsky_OpticalCoolingExperiment}.

\section{Nonlinear classical dynamic\label{sec:Nonlinear-classical-dynamic}}

Operating on the red detuned side of the resonance, any small thermal
oscillation amplitude will be damped away more quickly than in the
absence of radiation. On the opposite, blue detuned side, damping
is effectively reduced. If this effect overcomes intrinsic friction,
an arbitrary thermal fluctuation will be amplified into an oscillation
with increasing amplitude, driving the coupled system into a nonlinear
regime \cite{PhysRevA.36.3768,PhysRevA.49.1337,2001_Braginsky_ParametricInstability,2006_FM_DynamicalMultistability}.
Finally, the system will settle into a stable, self-sustained oscillation,
where radiation power input and dissipation are in balance. This will
be the subject of the present section. These effects have already
been observed in experiments \cite{2004_KarraiConstanze_IEEE,carmon:223902,kippenberg:033901,2008_Metzger}.

To derive classical equations of motion, we replace the operator $\hat{a}$
by the complex light amplitude $\alpha$ and the position operator
$\hat{x}_{M}$ by the cantilever's classical displacement $x_{M}$.
From the Hamiltonian Eqn.~(\ref{eq:Hamiltonian}) we then derive

\begin{eqnarray*}
\dot{\alpha} & = & \left[i\left(\Delta+g\,\frac{x_{M}}{x_{\mathrm{ZPF}}}\right)-\frac{\kappa}{2}\right]\alpha-i\alpha_{L}\\
\ddot{x}_{M} & = & -\omega_{M}^{2}x_{M}+\left|\alpha\right|^{2}\hbar g/(mx_{\mathrm{ZPF}})-\Gamma_{M}\dot{x}_{M}\,,\end{eqnarray*}
where $\alpha_{L}$ is the amplitude of the driving laser field, $\Gamma_{M}$
describes the mechanical damping of the cantilever, and $\Delta=\omega_{L}-\omega_{\mathrm{cav}}$
is the detuning of the laser light with respect to the cavity resonance. 

Beside a static solution $x_{M}(t)=\mathrm{const.}$, the system can
exhibit self-induced oscillations. The cantilever will then conduct
approximately sinusoidal oscillations, $x_{M}(t)\approx\bar{x}+A\cos(\omega_{M}t)$,
at its unperturbed frequency $\omega_{M}$. Since radiation pressure
effects are small, the amplitude $A$ of the oscillations will change
slowly over many oscillation periods only.

\begin{figure}
\begin{centering}
\includegraphics[width=0.6\columnwidth]{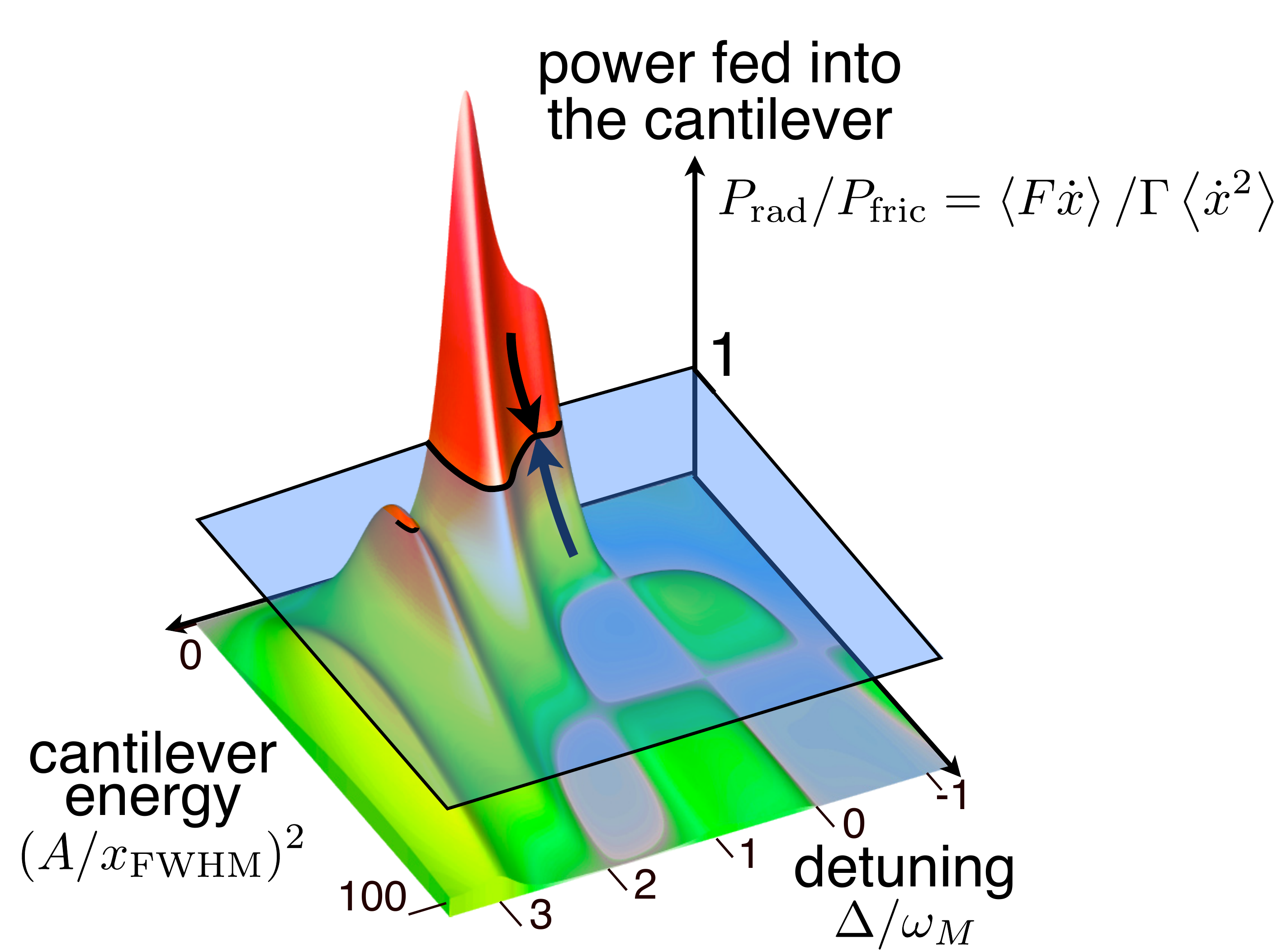}
\par\end{centering}

\caption{The power fed into the cantilever motion
by the radiation force, as a function of oscillation amplitude and
laser detuning. This can be used to construct the possible attractors
for the self-induced oscillations (indicated by thick lines).}
\label{fig:Classical-solution}
\end{figure}

From this ansatz, an analytical solution for the coupled dynamics
of $x_{M}(t)$ and $\alpha(t)$ can be found (\cite{2006_FM_DynamicalMultistability};
see also \cite{2008_ML_OptomechInstab}). The two parameters of the
solution, the amplitude $A$ and the average displacement $\bar{x}$,
can be determined from two balance conditions: For any periodic solution
the total force should average to zero during one cycle,\begin{equation}
\left\langle \ddot{x}_{M}\right\rangle \equiv0\quad\Leftrightarrow\quad m\omega_{M}^{2}\bar{x}=\left\langle F_{\mathrm{rad}}\right\rangle =\frac{\hbar g}{mx_{\mathrm{ZPF}}}\left\langle \left|\alpha(t)\right|^{2}\right\rangle \,.\label{eq:force balance}\end{equation}
This yields an implicit equation for $\bar{x}$, since $\left\langle F_{\mathrm{rad}}\right\rangle $
is a function of $\bar{x}$ and $A$. Furthermore, the work performed
by the radiation pressure balances on average the frictional losses,\begin{equation}
\left\langle F_{\mathrm{rad}}\dot{x}\right\rangle =\Gamma_{M}\left\langle \dot{x}^{2}\right\rangle \,.\label{eq:power balance}\end{equation}
Eliminating $\bar{x}$ by use of Eqn.~(\ref{eq:force balance}) we
can plot the ratio between radiation power input and frictional loss,
the two sides of the last equation, as a function of the oscillation
amplitude $A$. Such a plot is shown in Fig.~\ref{fig:Classical-solution},
where we chose the detuning $\Delta$ as a second variable, while
other parameters are fixed. The condition of Eqn.~(\ref{eq:power balance})
is fulfilled if the ratio $P_{\mathrm{rad}}/P_{\mathrm{fric}}=1$,
as indicated by the horizontal cut in Fig.~\ref{fig:Classical-solution}.
A solution will be stable only if an increase of the amplitude is
accompanied by a decrease of $P_{\mathrm{rad}}/P_{\mathrm{fric}}$.
By that reasoning the final attractor diagram is constructed, as indicated
by the thick black lines in Fig.~\ref{fig:Classical-solution}. 

Important general features of the dynamics of the coupled system can
be seen in Fig.~\ref{fig:Classical-solution}. Self-induced oscillations
appear for sufficiently strong driving around integer multiples of
the cantilever frequency, $\Delta\approx n\omega_{M}$. Such oscillations
appear for a positive detuning $\Delta$, while for red detuned laser
light ($\Delta<0$) the stationary solution, $x_{M}(t)=\mathrm{const.}$,
is stable. Note that stable solutions with large amplitude do exist
even for $\Delta<0$.

The most striking feature, however, is the coexistence of several
stable solutions with different finite oscillation amplitudes for
a fixed set of system parameters. This dynamical multi-stability,
first discussed in this context in Ref.~\cite{2006_FM_DynamicalMultistability}
and also seen in similar systems \cite{2007_Rodrigues_DynInstabilitiesSSET},
is visible in Fig.~\ref{fig:Classical-solution}(b), while for the
parameters of Fig.~\ref{fig:Classical-solution}(a) we find coexistence
of a stationary and a finite amplitude solution around $\Delta\approx2\omega_{M}$.

These multi-stabilities could be utilized for ultra-sensitive {}``latching''
measurements, as argued in Ref.~\cite{2006_FM_DynamicalMultistability}.

Self-induced oscillations in an optomechanical system have already
been observed in experiments with bolometric forces \cite{2004_KarraiConstanze_IEEE,2008_Metzger}
and in microtoroidal structures where radiation pressure dominates
\cite{carmon:223902}. Recently, a more detailed comparison of theory
and experiment revealed interesting effects due to higher order mechanical
modes that get involved in the nonlinear dynamics \cite{2008_Metzger}.

\section{Quantum theory of optomechanical systems\label{sec:Quantum-theory-of}}

The prospect of reaching the quantum mechanical ground state of a
{}``macroscopic'' mechanical object is currently one of the main
goals in the field of micro- and nanomechanics. Impressive progress
has been made in a series of experiments \cite{PhysRevLett.83.3174,2004_12_ConstanzeKhaled_WithNote,2006_07_Arcizet_CoolingMirror,2006_Gigan_Self-cooling,schliesser:243905,2006_11_Bouwmeester_FeedbackCooling,2007_Corbitt_MirrorTrap,Thompson_2008},
though the ground state has not yet been reached at the time of writing.
In the classical picture derived above, we found that a properly detuned
laser beam will cool the cantilever by providing extra damping. According
to the classical theory, the cantilever can be cooled down to an effective
temperature $T_{\mathrm{eff}}=T\,\Gamma_{M}/(\Gamma_{\mathrm{opt}}+\Gamma_{M})$,
apparently arbitrarily close to absolute zero for sufficient drive
power and low mechanical damping. However, quantum mechanics sets
the ultimate limit for optomechanical cooling.

Starting from an intuitive quantum picture of the cooling process,
we will present in the next subsection a quantum noise approach to
cooling. Quantum effects on the self-induced oscillations can be described
numerically within a quantum master equation discussed in the following
subsection, which allows studying the classical-to-quantum crossover.

\subsection{Quantum noise approach to cooling}

In the quantum description, 
a photon impinging on the cavity will emit or absorb a phonon of the
mechanical cantilever motion and change its frequency accordingly, in a Raman-like process.
A photon that is red detuned from the resonance will absorb a phonon
of energy $\hbar\omega_{M}$ from the cantilever, so that it is scattered
into the cavity resonance, leading to cooling. Detuning to a `sideband'
of the cavity at a frequency $\omega_{\mathrm{cav}}-\omega_{M}$ will
be particularly effective.

\begin{figure}
\includegraphics[width=1\columnwidth]{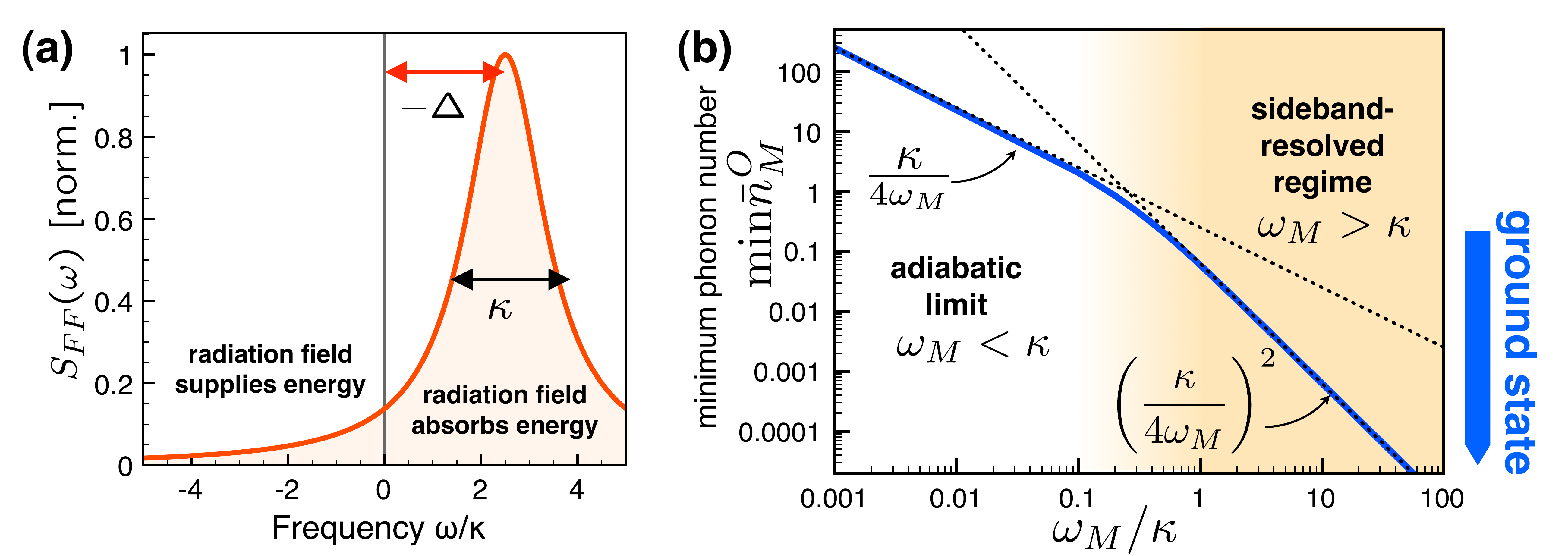}

\caption{(a) Power spectrum for the radiation pressure
force. (b) Quantum-mechanical cooling limit.}
\label{fig:Power-spectrum}
\end{figure}

For a quantitative approach the radiation field of the cavity will
be considered as a `bath' acting upon the `system', the cantilever
degree of freedom $\hat{x}_{M}$, via the coupling term, $-\hat{x}_{M}\hat{F}$,
in the Hamiltonian. The influence of the bath is then characterized
by the power spectrum of the force, $\hat{S}_{FF}(\omega)=\int dt\,\exp(i\omega t)$$\left\langle \hat{F}(t)\hat{F}(0)\right\rangle $
. In particular, Fermi's golden rule links the net optical damping
rate of the cantilever to the possibility of the cavity to absorb/emit
a quantum of energy $\hbar\omega_{M}$ from/to the bath, $S_{FF}(\pm\omega_{M})$,
as\begin{equation}
\Gamma_{\mathrm{opt}}=(x_{\mathrm{ZPF}}/\hbar)^{2}\left[S_{FF}(\omega_{M})-S_{FF}(-\omega_{M})\right]\,.\label{eq:optical damping rate}\end{equation}
The power spectrum $S_{FF}$ is directly related \cite{2007_FM_SidebandCooling}
to the spectrum of photon number fluctuations due to shot-noise (see
Fig.~\ref{fig:Power-spectrum}). Crucially, the asymmetry of the
power spectrum (which is set by the laser detuning) determines whether
the cavity will more readily absorb or emit energy, setting the sign
of the net optical damping rate $\Gamma_{\mathrm{opt}}$ {[}cf. Eqn.~(\ref{eq:optical damping rate}){]}.

One finds \cite{2007_FM_SidebandCooling,wilson-rae:093901} a simple
limit on the minimal occupation number, $\bar{n}_{M}^{O}=[\kappa/(4\omega_{M})]^{2}$,
which can be reached for optimal detuning $\Delta=-\omega_{M}$ in
the resolved-sideband limit $\omega_{M}\gg\kappa$, for $\Gamma_{\mathrm{opt}}\gg\Gamma_{M}$.
In general, the reachable occupation number $\bar{n}_{M}$ of the
mechanical mode will depend on the initial occupation $\bar{n}_{M}^{T}$
(hence, starting from cryogenically precooled samples is advantageous)
and the mechanical and optical damping rates, as $\bar{n}_{M}=(\Gamma_{\mathrm{opt}}\bar{n}_{M}^{O}+\Gamma_{M}\bar{n}_{M}^{T})/(\Gamma_{\mathrm{opt}}+\Gamma_{M})\,,$
which reduces to the simple classical expression for the effective
temperature given above for $\bar{n}_{M}^{T}\gg1$. As shown in Fig.~\ref{fig:Power-spectrum}
ground state cooling is most advantageously pursued in the resolved-sideband
regime with high finesse cavities and high frequency resonators. With
various groups working on a variety of setups further progress and
final success in approaching the quantum limit is expected in the
very near future.

The strong coupling regime, where $\Gamma_{{\rm opt}}>\kappa$, needs
a more sophisticated analysis and gives rise to new features \cite{2007_FM_SidebandCooling,2008_FM_Cooling_Review}.

\subsection{Quantum description of self-induced oscillations\label{sub:Quantum-description-of}}

For a full quantum description \cite{2008_ML_OptomechInstab} of the
self-induced oscillations, we have to consider the reduced density
matrix $\hat{\rho}$ of the system consisting of cantilever and cavity
mode. Mechanical damping and photon decay out of the cavity are treated
using a Lindblad master equation,\begin{equation}
\frac{d}{dt}\hat{\rho}=\mathcal{L}\hat{\rho}=-\frac{i}{\hbar}\left[\hat{H}_{\mathrm{cav+cant+drive}},\,\hat{\rho}\right]+\Gamma_{M}\mathcal{D}\left[\hat{c}\right]+\kappa\mathcal{D}\left[\hat{a}\right]\quad\mathrm{(for}\quad T=0\mathrm{)},\label{eq:master}\end{equation}
where $\mathcal{D}\left[\hat{a}\right]=\hat{a}\hat{\rho}\hat{a}^{\dagger}-\frac{1}{2}\hat{a}^{\dagger}\hat{a}\hat{\rho}-\frac{1}{2}\hat{\rho}\hat{a}^{\dagger}\hat{a}$
is of the standard Lindblad form.

The stationary state of the system is found as the eigenvector of
the Liouvillian $\mathcal{L}$ for eigenvalue zero. This problem can
be solved numerically for a restricted, but sufficiently large number
of cavity and cantilever states. From the eigenvector, the density
matrix $\hat{\rho}_{f}$, all quantities of interest, for instance,
the average kinetic energy of the cantilever motion, can then be calculated.

Before comparing the results of this quantum mechanical description
to the classical approach, it is instructive to quantify the degree
of `quantumness' of the system. Using the dimensionless parameters
$\mathcal{P}=8\left|\alpha_{L}\right|^{2}g^{2}/\omega_{M}^{4}$, characterizing
the driving strength, and $\zeta=g/\kappa$, the Hamiltonian is written
as\begin{equation}
\hat{H}_{\mathrm{cav+cant+drive}}=\hbar\left\{ \left[-\Delta-\kappa\zeta(\hat{c}+\hat{c}^{\dagger})\right]\hat{a}^{\dagger}\hat{a}+\omega_{M}\hat{c}^{\dagger}\hat{c}+\frac{\sqrt{2\mathcal{P}}\omega_{M}^{2}}{4\kappa\zeta}(\hat{a}+\hat{a}^{\dagger})\right\} \,.\label{eq:coherent part}\end{equation}
The master Eqn.~(\ref{eq:master}) then contains only dimensionless
quantities, if time and the remaining energy/frequency variables are
written in terms of the mechanical oscillation frequency $\omega_{M}$.
Four of the dimensionless parameters in this equation, $\Gamma_{M}/\omega_{M},\,\kappa/\omega_{M},\,\Delta/\omega_{M}$
and $\mathcal{P}$ do also appear in the classical equations of motion,
while\begin{equation}
\zeta=\frac{g}{\kappa}=\frac{x_{\mathrm{ZPF}}}{x_{\mathrm{FWHM}}}\propto\sqrt{\hbar}\label{eq:quantum parameter}\end{equation}
does not. The so-defined `quantum parameter' $\zeta$ constitutes
a measure of the quantum nature of the system and vanishes in the
classical limit $\hbar\rightarrow0$. It is defined as the ratio of
the quantum mechanical zero point fluctuations of the cantilever to
a classical length scale, namely the resonance width $x_{\mathrm{FWHM}}$
of the cavity.

The quantum master equation allows studying the quantum-to-classical
crossover of the system dynamics by changing the numerical value of
the quantum parameter $\zeta$. Classical results are recovered for
small $\zeta$, while for $\zeta\gtrsim1$ quantum fluctuations tend
to smear out the sharp features of the classical result and favour
the occurrence of self-induced oscillations below the classical onset,
a feature which can also be deduced from the quantum noise approach
introduced above (see Ref.~\cite{2008_ML_OptomechInstab} for details
and figures). Note that to some extent the effects of quantum fluctuations
can be mimicked by introducing quantum zero-point fluctuations into
the classical equations of motion \cite{2008_ML_OptomechInstab}. 

\begin{figure}
\begin{centering}
\includegraphics[width=0.8\columnwidth]{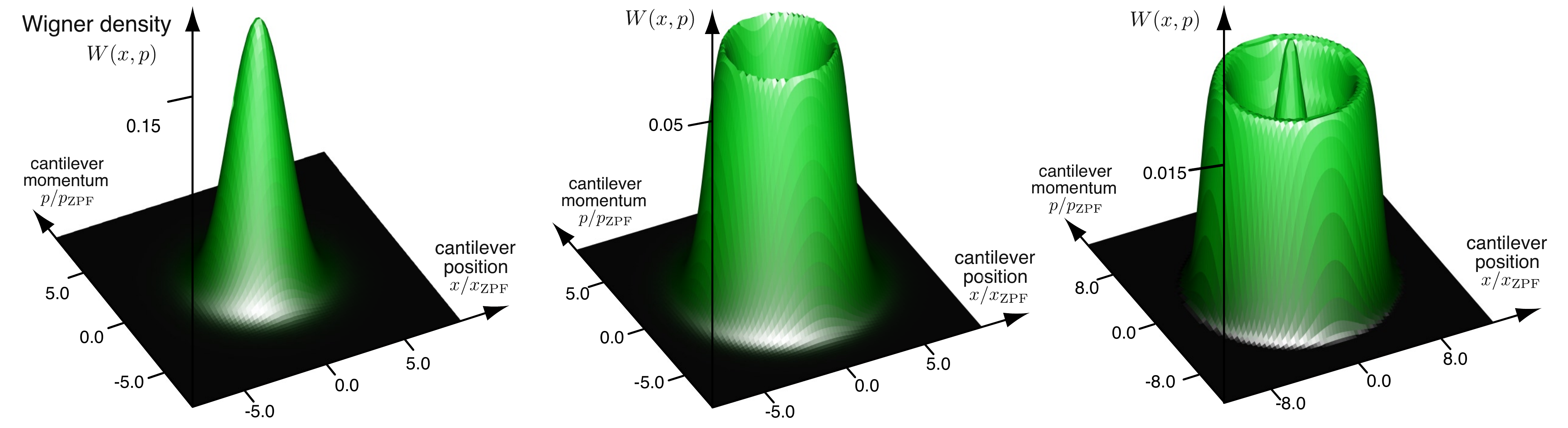}
\par\end{centering}

\caption{Wigner densities for the cantilever under
the influence of the radiation force, for varying detuning, displaying the optomechanical
instability (middle and right panels).}
\label{fig:Wigner-densities}
\end{figure}

The existence of classical bi- or multistable solutions can be seen
by considering the Wigner density of the cantilever. As illustrated
in Fig.~\ref{fig:Wigner-densities}, the Wigner density shows characteristic
features corresponding to (a) a single stationary classical solution
(broad peak in phase space), (b) a single finite amplitude classical
solution (ring structure - the phase of the oscillatory solution is
undetermined), or (c) the coexistence of a classical stationary and
finite amplitude solution (peak with superimposed ring structure). 

We find that most optomechanical experiments are well in the classical
regime, in the sense that the quantum parameter remains small (e.g.,
$\zeta\approx10^{-3}\ll1$ in the Bouwmeester setup \cite{2006_11_Bouwmeester_FeedbackCooling}).
In two recent setups, however, combining standard optomechanics with
cold-atom physics \cite{2007_Gupta,2008_Murch_Observation_nature,2008_Brennecke_CavityOptomechanics},
$\zeta$ is of the order of one. In these experiments a cloud of cold
atoms is placed in an optical cavity, so that the collective motion
of the cloud couples to an optical mode of the cavity, replacing the
cantilever motion.

\section{Towards Fock-state detection\label{sec:Towards-Fock-state-detection}}

Linked inextricably to the race towards ground state cooling is the
question how to confirm the quantum nature of the final state. Measurement
of the displacement quadratures is possible via optical readout \cite{2008_02_Clerk_Squeezing}. However, probably
the most straightforward demonstration would be to observe the quantum
jumps from the ground state to progressively higher energy eigenstates
(Fock states), as the system heats up again. Such quantum jumps between
different Fock states have been observed in the mechanical motion
of an electron in a Penning trap \cite{peil:1287}. In optomechanics
such quantum jumps might eventually be observed for the mechanical
motion of a truly macroscopic object, consisting of billions of atoms.

Recently, the Yale group of Jack Harris introduced a novel optomechanical
setup \cite{Thompson_2008,2008_Jayich_NJP}, where a thin dielectric
membrane is placed in the middle of a cavity with two fixed, high
finesse mirrors. Beside the technological advances offered by this
setup, it also leads to a different coupling of the mechanical displacement
of the oscillating membrane to the cavity, which is advantageous for
the aim of Fock state detection. To find the structure of the coupling
term in the Hamiltonian, consider first the limit of a perfectly reflecting
membrane at some position $x$ in the middle of the cavity. Moving
the membrane will change the frequencies of resonances in the left
and right halves of the cavity in opposite directions, which would
lead to a resonance crossing at some displacement $x_{\mathrm{cross}}$.
A finite transmission of the membrane, however, produces an anti-crossing,
with $\omega(x)-\omega(x_{\mathrm{cross}})\propto x^{2}$ near the
degeneracy point. In rotating wave approximation the coupling is then
of the form $\propto(\hat{c}^{\dagger}\hat{c}+\frac{1}{2})\,\hat{a}^{\dagger}\hat{a}$,
so that $\left[\hat{H}_{\mathrm{cant+cav+drive}},\,\hat{c}^{\dagger}\hat{c}\right]=0$,
allowing non-destructive measurement of the phonon number. Detecting
the phase of the transmitted beam driving the cavity at resonance
frequency then constitutes a direct quantum non-demolition (QND) measurement
of the phonon number.

Shot noise in the transmitted beam can be overcome by time averaging,
which, however, is restricted by the life time of Fock states due
to finite damping and temperature. Optimal averaging times and strategies,
how best to distinguish classical from quantum fluctuations, even
when the QND readout time is comparable to the state's life time,
have been explored in Ref.~\cite{2008_Jayich_NJP}.

\section{Conclusions}

Optomechanics is a new research topic that has been established in
the past four years, with strong progress being made through a tight
interplay of theory and experiment. Even the classical nonlinear dynamics
of these systems is far from being fully explored: For example, chaotic
motion has been observed at strong drive \cite{carmon:223902},
but not yet analyzed systematically. In the quantum regime, ground-state
cooling and creation of nonclassical states (e.g. entanglement) are
interesting challenges. New setups expand the applicability of these
concepts, e.g. in superconducting microwave circuits or with cold
atoms.

F.M. acknowledges the inspiring atmosphere of the NATO workshop in
Tashkent, Usbekistan, where this overview was presented, as well as 
J. Harris, S. Girvin, K. Karrai, C. Metzger, A. Clerk, and C. Neuenhahn for
collaboration on this topic. This research
is funded by the German Science Foundation (DFG) through NIM, SFB631,
and the Emmy-Noether program.

\bibliographystyle{nato}
\bibliography{bib_Max_FM_comb}

\begin{thebibliography}{}

\bibitem[\protect\citeauthoryear{Aguirregabiria and
  Bel}{1987}]{PhysRevA.36.3768}
Aguirregabiria, J.~M. and Bel, L. (1987) Delay-induced instability in a
  pendular Fabry-Perot cavity,
\newblock {\em Phys. Rev. A} {\bf 36}, 3768--3770.

\bibitem[\protect\citeauthoryear{Arcizet
  et~al.}{2006}]{2006_07_Arcizet_CoolingMirror}
Arcizet, O., Cohadon, P.~F., Briant, T., Pinard, M., and Heidmann, A. (2006)
  Radiation-pressure cooling and optomechanical instability of a micro-mirror,
\newblock {\em Nature} {\bf 444}, 71.

\bibitem[\protect\citeauthoryear{Braginsky and
  Manukin}{1967}]{1967_BraginskyManukin_PonderomotiveEffectsEMRadiation}
Braginsky, V. and Manukin, A. (1967) Ponderomotive effects of electromagnetic
  radiation,
\newblock {\em Soviet Physics JETP} {\bf 25}, 653.

\bibitem[\protect\citeauthoryear{Braginsky
  et~al.}{1970}]{1970_Braginsky_OpticalCoolingExperiment}
Braginsky, V.~B., Manukin, A.~B., and Tikhonov, M.~Y. (1970) Investigation of
  dissipative ponderomotove effects of electromagnetic radiation,
\newblock {\em Soviet Physics JETP} {\bf 31}, 829.

\bibitem[\protect\citeauthoryear{Braginsky
  et~al.}{2001}]{2001_Braginsky_ParametricInstability}
Braginsky, V.~B., Strigin, S.~E., and Vyatchanin, S.~P. (2001) Parametric
  oscillatory instability in Fabry-Perot interferometer,
\newblock {\em Physics Letters A} {\bf 287}, 331--338.

\bibitem[\protect\citeauthoryear{Brennecke
  et~al.}{2008}]{2008_Brennecke_CavityOptomechanics}
Brennecke, F., Ritter, S., Donner, T., and Esslinger, T. (2008) Cavity
  Optomechanics with a Bose-Einstein Condensate,
\newblock {\em Science} {\bf 322}, 235--238.

\bibitem[\protect\citeauthoryear{Brown et~al.}{2007}]{brown:137205}
Brown, K.~R., Britton, J., Epstein, R.~J., Chiaverini, J., Leibfried, D., and
  Wineland, D.~J. (2007) Passive Cooling of a Micromechanical Oscillator with a
  Resonant Electric Circuit,
\newblock {\em Physical Review Letters} {\bf 99}, 137205.

\bibitem[\protect\citeauthoryear{Carmon et~al.}{2005}]{carmon:223902}
Carmon, T., Rokhsari, H., Yang, L., Kippenberg, T.~J., and Vahala, K.~J. (2005)
  Temporal Behavior of Radiation-Pressure-Induced Vibrations of an Optical
  Microcavity Phonon Mode,
\newblock {\em Physical Review Letters} {\bf 94}, 223902.

\bibitem[\protect\citeauthoryear{Clerk et~al.}{2008}]{2008_02_Clerk_Squeezing}
Clerk, A.~A., Marquardt, F., and Jacobs, K. (2008) Back-action evasion and
  squeezing of a mechanical resonator using a cavity detector,
\newblock {\em New Journal of Physics} {\bf 10}, 095010.

\bibitem[\protect\citeauthoryear{Cohadon et~al.}{1999}]{PhysRevLett.83.3174}
Cohadon, P.~F., Heidmann, A., and Pinard, M. (1999) Cooling of a Mirror by
  Radiation Pressure,
\newblock {\em Phys. Rev. Lett.} {\bf 83}, 3174--3177.

\bibitem[\protect\citeauthoryear{Corbitt
  et~al.}{2007}]{2007_Corbitt_MirrorTrap}
Corbitt, T., Chen, Y., Innerhofer, E., Muller-Ebhardt, H., Ottaway, D.,
  Rehbein, H., Sigg, D., Whitcomb, S., Wipf, C., and Mavalvala, N. (2007) An
  All-Optical Trap for a Gram-Scale Mirror,
\newblock {\em Physical Review Letters} {\bf 98}, 150802.

\bibitem[\protect\citeauthoryear{D~A~Rodrigues and
  Armour}{2007}]{2007_Rodrigues_DynInstabilitiesSSET}
D~A~Rodrigues, J~Imbers, T. J.~H. and Armour, A.~D. (2007) Dynamical
  instabilities of a resonator driven by a superconducting single-electron
  transistor,
\newblock {\em New Journal of Physics} {\bf 9}, 84.

\bibitem[\protect\citeauthoryear{Fabre et~al.}{1994}]{PhysRevA.49.1337}
Fabre, C., Pinard, M., Bourzeix, S., Heidmann, A., Giacobino, E., and Reynaud,
  S. (1994) Quantum-noise reduction using a cavity with a movable mirror,
\newblock {\em Phys. Rev. A} {\bf 49}, 1337--1343.

\bibitem[\protect\citeauthoryear{Gigan et~al.}{2006}]{2006_Gigan_Self-cooling}
Gigan, S., Bohm, H.~R., Paternostro, M., Blaser, F., Langer, G., Hertzberg,
  J.~B., Schwab, K.~C., Bauerle, D., Aspelmeyer, M., and Zeilinger, A. (2006)
  Self-cooling of a micromirror by radiation pressure,
\newblock {\em Nature} {\bf 444}, 67--70.

\bibitem[\protect\citeauthoryear{Gupta et~al.}{2007}]{2007_Gupta}
Gupta, S., Moore, K.~L., Murch, K.~W., and Stamper-Kurn, D.~M. (2007) Cavity
  Nonlinear Optics at Low Photon Numbers from Collective Atomic Motion,
\newblock {\em Physical Review Letters} {\bf 99}, 213601.

\bibitem[\protect\citeauthoryear{{H\"ohberger} and
  Karrai}{2004}]{2004_KarraiConstanze_IEEE}
{H\"ohberger}, C. and Karrai, K. (2004) Self-oscillation of micromechanical
  resonators,
\newblock {\em Nanotechnology 2004, Proceedings of the 4th IEEE conference on
  nanotechnology} p. 419.

\bibitem[\protect\citeauthoryear{{H\"ohberger}-Metzger and
  Karrai}{2004}]{2004_12_ConstanzeKhaled_WithNote}
{H\"ohberger}-Metzger, C. and Karrai, K. (2004) Cavity cooling of a microlever,
\newblock {\em Nature} {\bf 432}, 1002.

\bibitem[\protect\citeauthoryear{Jayich et~al.}{2008}]{2008_Jayich_NJP}
Jayich, A.~M., Sankey, J.~C., Zwickl, B.~M., Yang, C., Thompson, J.~D., Girvin,
  S.~M., Clerk, A.~A., Marquardt, F., and Harris, J. G.~E. (2008) Dispersive
  optomechanics: a membrane inside a cavity,
\newblock {\em New Journal of Physics} {\bf 10}, 095008 (28pp).

\bibitem[\protect\citeauthoryear{Kippenberg et~al.}{2005}]{kippenberg:033901}
Kippenberg, T.~J., Rokhsari, H., Carmon, T., Scherer, A., and Vahala, K.~J.
  (2005) Analysis of Radiation-Pressure Induced Mechanical Oscillation of an
  Optical Microcavity,
\newblock {\em Physical Review Letters} {\bf 95}, 033901.

\bibitem[\protect\citeauthoryear{Kippenberg and
  Vahala}{2008}]{2008_KippenbergVahala_ScienceReview}
Kippenberg, T.~J. and Vahala, K.~J. (2008) Cavity Optomechanics: Back-Action at
  the Mesoscale,
\newblock {\em Science} {\bf 321}, 5893.

\bibitem[\protect\citeauthoryear{Kleckner and
  Bouwmeester}{2006}]{2006_11_Bouwmeester_FeedbackCooling}
Kleckner, D. and Bouwmeester, D. (2006) Sub-kelvin optical cooling of a
  micromechanical resonator,
\newblock {\em Nature} {\bf 444}, 75.

\bibitem[\protect\citeauthoryear{Ludwig et~al.}{2008}]{2008_ML_OptomechInstab}
Ludwig, M., Kubala, B., and Marquardt, F. (2008) The optomechanical instability
  in the quantum regime,
\newblock {\em New Journal of Physics} {\bf 10}, 095013 (19pp).

\bibitem[\protect\citeauthoryear{Marquardt}{2008}]{2008_FM_NaturePhysNewsAndVi%
ews}
Marquardt, F. (2008) Optomechanics: Push towards the quantum limit,
\newblock {\em Nat Phys} {\bf 4}, 513--514.

\bibitem[\protect\citeauthoryear{Marquardt
  et~al.}{2007}]{2007_FM_SidebandCooling}
Marquardt, F., Chen, J.~P., Clerk, A.~A., and Girvin, S.~M. (2007) Quantum
  Theory of Cavity-Assisted Sideband Cooling of Mechanical Motion,
\newblock {\em Physical Review Letters} {\bf 99}, 093902.

\bibitem[\protect\citeauthoryear{Marquardt
  et~al.}{2008}]{2008_FM_Cooling_Review}
Marquardt, F., Clerk, A.~A., and Girvin, S.~M. (2008) Quantum theory of
  optomechanical cooling,
\newblock {\em Journal of Modern Optics} {\bf 55}, 3329.

\bibitem[\protect\citeauthoryear{Marquardt
  et~al.}{2006}]{2006_FM_DynamicalMultistability}
Marquardt, F., Harris, J. G.~E., and Girvin, S.~M. (2006) Dynamical
  Multistability Induced by Radiation Pressure in High-Finesse Micromechanical
  Optical Cavities,
\newblock {\em Physical Review Letters} {\bf 96}, 103901.

\bibitem[\protect\citeauthoryear{Metzger et~al.}{2008}]{2008_Metzger}
Metzger, C., Ludwig, M., Neuenhahn, C., Ortlieb, A., Favero, I., Karrai, K.,
  and Marquardt, F. (2008) Self-Induced Oscillations in an Optomechanical
  System Driven by Bolometric Backaction,
\newblock {\em Physical Review Letters} {\bf 101}, 133903.

\bibitem[\protect\citeauthoryear{Murch
  et~al.}{2008}]{2008_Murch_Observation_nature}
Murch, K.~W., Moore, K.~L., Gupta, S., and Stamper-Kurn, D.~M. (2008)
  Observation of quantum-measurement backaction with an ultracold atomic gas,
\newblock {\em Nat Phys} {\bf 4}, 561--564.

\bibitem[\protect\citeauthoryear{Naik
  et~al.}{2006}]{2006_Naik_CoolingNanomechResonator}
Naik, A., Buu, O., LaHaye, M.~D., Armour, A.~D., Clerk, A.~A., Blencowe, M.~P.,
  and Schwab, K.~C. (2006) Cooling a nanomechanical resonator with quantum
  back-action,
\newblock {\em Nature} {\bf 443}, 193--196.

\bibitem[\protect\citeauthoryear{Peil and Gabrielse}{1999}]{peil:1287}
Peil, S. and Gabrielse, G. (1999) Observing the Quantum Limit of an Electron
  Cyclotron: QND Measurements of Quantum Jumps between Fock States,
\newblock {\em Physical Review Letters} {\bf 83}, 1287--1290.

\bibitem[\protect\citeauthoryear{Regal
  et~al.}{2008}]{2008_Regal_MicrowaveCavity}
Regal, C.~A., Teufel, J.~D., and Lehnert, K.~W. (2008) Measuring nanomechanical
  motion with a microwave cavity interferometer,
\newblock {\em Nat Phys} {\bf 4}, 555--560.

\bibitem[\protect\citeauthoryear{Schliesser et~al.}{2006}]{schliesser:243905}
Schliesser, A., Del'Haye, P., Nooshi, N., Vahala, K.~J., and Kippenberg, T.~J.
  (2006) Radiation Pressure Cooling of a Micromechanical Oscillator Using
  Dynamical Backaction,
\newblock {\em Physical Review Letters} {\bf 97}, 243905.

\bibitem[\protect\citeauthoryear{Schwab and
  Roukes}{2005}]{2005_SchwabRoukes_PhysicsToday}
Schwab, K.~C. and Roukes, M.~L. (2005) Putting Mechanics into Quantum
  Mechanics,
\newblock {\em Physics Today} {\bf July}, 36.

\bibitem[\protect\citeauthoryear{Thompson et~al.}{2008}]{Thompson_2008}
Thompson, J.~D., Zwickl, B.~M., Jayich, A.~M., Marquardt, F., Girvin, S.~M.,
  and Harris, J. G.~E. (2008) Strong dispersive coupling of a high-finesse
  cavity to a micromechanical membrane,
\newblock {\em Nature} {\bf 452}, 900--900.

\bibitem[\protect\citeauthoryear{Wilson-Rae et~al.}{2007}]{wilson-rae:093901}
Wilson-Rae, I., Nooshi, N., Zwerger, W., and Kippenberg, T.~J. (2007) Theory of
  Ground State Cooling of a Mechanical Oscillator Using Dynamical Backaction,
\newblock {\em Physical Review Letters} {\bf 99}, 093901.

\end{thebibliography}

\end{document}